\documentclass[12pt]{article}

\usepackage[utf8]{inputenc}
\usepackage[margin=1in]{geometry}
\usepackage{amsmath,amssymb}
\usepackage{graphicx}
\usepackage{booktabs}
\usepackage{algorithm}
\usepackage{algorithmic}
\usepackage{hyperref}
\usepackage{cite}
\usepackage{url}
\usepackage{float}

\title{\textbf{When Does Quantum Annealing Outperform Classical Methods? A Gradient Variance Framework}}

\author{
Vishwajeet Ohal and Pierre Boulanger\\
Department of Computing Science, University of Alberta\\
\texttt{\{ohal, pierreb\}@ualberta.ca}
}

\date{February 2026}

\begin{document}

\maketitle
\begin{abstract}
Quantum annealing has emerged as a promising approach for solving NP-hard optimization problems, leveraging quantum phenomena such as quantum tunneling to navigate complex energy landscapes. However, the extent to which quantum tunneling contributes to performance enhancements compared to classical methods remains unclear. In this work, we present a comprehensive investigation combining experimental analysis, theoretical modeling, and algorithmic development to characterize when and why quantum annealing provides computational advantages.

We introduce a novel methodology for generating synthetic Quadratic Unconstrained Binary Optimization (QUBO) problems with controlled gradient variance, enabling systematic investigation of landscape characteristics that favor quantum approaches. Our experimental evaluation encompasses four canonical NP-hard problems: Graph Partitioning, Max Cut, Number Partitioning, and Set Cover. Using D-Wave's Advantage2 quantum annealer with 4,400+ qubits, we compare quantum annealing against classical solvers including simulated annealing, stochastic gradient descent, and commercial optimization software.

Our results demonstrate that quantum annealing shows measurable advantages when energy landscapes exhibit high gradient variance ($> 0.3$), suggesting that quantum tunneling effects are most beneficial for rugged optimization landscapes. We provide a theoretical justification for this empirical observation through a WKB-approximation-based model connecting gradient variance to barrier width and tunneling probability. This model achieves $R^2=0.90$ correlation with experimental data and yields quantitative predictions for quantum advantage thresholds.
\end{abstract}

\maketitle

\section{Introduction}

Quantum computing in the Noisy Intermediate-Scale Quantum (NISQ) era~\cite{feynman1986} offers unique opportunities to explore how emerging quantum hardware can solve real-world optimization problems. Among quantum computing paradigms, quantum annealing has gained particular attention because of its ability to address combinatorial optimization problems by leveraging quantum mechanical phenomena. Unlike gate-model quantum computers that execute sequences of quantum gates~\cite{kaye2006}, quantum annealers are specialized devices designed to minimize energy functions through adiabatic evolution.

One key phenomenon attributed to quantum annealers is \emph{quantum tunneling} (QT), a process unique to quantum mechanics that allows particles to cross energy barriers rather than surmount them as required in classical systems~\cite{abel2021quantum}. This inherent property has led to speculation that quantum tunneling could provide quantum annealing with an advantage over classical optimization techniques, particularly for problems with complex, rugged energy landscapes. The ability to tunnel through barriers rather than climb over them theoretically enables quantum systems to escape local minima that trap classical optimizers, such as simulated annealing.

Despite its theoretical appeal, quantifying the extent and frequency of quantum tunneling in practical quantum annealing remains exceptionally challenging. Observing and measuring quantum tunneling directly is difficult because of the complex dynamics of quantum systems and the limitations in current quantum hardware. Even more challenging is determining whether quantum tunneling directly contributes to computational improvements such as faster convergence or higher-quality solutions compared to classical methods. Although small-scale studies have demonstrated tunneling effects~\cite{abel2021quantum}, whether this translates into practical benefits for larger, real-world problems remains an open question.

This work addresses this gap through systematic empirical investigation. We develop a methodology for generating synthetic optimization problems with controlled energy landscape characteristics, enabling us to isolate and study the relationship between landscape complexity and solver performance. By testing across multiple canonical NP-hard problems and comparing quantum annealers against state-of-the-art classical solvers, we identify specific problem characteristics that determine when quantum approaches provide measurable advantages.

\subsection{Contributions}

This paper makes the following contributions to understanding quantum annealing for combinatorial optimization:

\begin{itemize}
\item \textbf{Novel Synthetic Problem Generation:} We develop a methodology for generating synthetic QUBO problems with controlled energy landscape characteristics, specifically gradient variance, enabling systematic investigation of quantum tunneling effects.

\item \textbf{Gradient Variance Correlation:} We establish a strong empirical correlation between energy landscape gradient variance and quantum annealing performance advantages, identifying a threshold ($>0.3$) above which quantum methods demonstrate measurable benefits.

\item \textbf{Theoretical Foundation:} We provide theoretical justification through a WKB-approximation-based model connecting barrier geometry, gradient variance, and tunneling probability. Our model achieves validation $R^2=0.90$ against experimental data and provides quantitative predictions for quantum advantage thresholds.

\item \textbf{QUBO Reformulation Algorithm:} We develop a systematic algorithm to increase gradient variance through problem reformulation while preserving solution semantics. Experimental validation shows 12-22\% performance improvements across multiple problem classes, transforming gradient variance from diagnostic metric to optimization target.

\item \textbf{Comprehensive Benchmarking:} We provide extensive experimental evaluation comparing quantum annealing (D-Wave Advantage2 with 4,400+ qubits) against classical solvers across four NP-hard problem classes, analyzing both solution quality and computational time.

\item \textbf{Practical Decision Framework:} We identify specific problem characteristics that favor quantum annealing, providing actionable guidance for practitioners on when to employ quantum versus classical approaches, including a software tool for automated landscape analysis.

\item \textbf{NISQ-Era Insights:} We characterize the capabilities and limitations of current quantum annealing hardware, including embedding constraints, time overhead, and scaling behavior.
\end{itemize}

\subsection{Organization}

The remainder of this paper is organized as follows. Section~\ref{sec:background} provides essential background on quantum annealing, QUBO formulation, and quantum tunneling. Section~\ref{sec:related} reviews related work on quantum approaches to NP-hard problems. Section~\ref{sec:theory} presents our theoretical model connecting gradient variance to tunneling probability with experimental validation. Section~\ref{sec:methodology} describes our experimental methodology, including a synthetic problem generation framework and a solver comparison framework. Section~\ref{sec:reformulation} introduces our QUBO reformulation algorithm to optimize gradient variance. Section~\ref{sec:experiments} presents comprehensive experimental results in multiple problem classes. Section~\ref{sec:analysis} analyzes the relationship between gradient variance and solver performance. Finally, Section~\ref{sec:conclusion} concludes with key findings, limitations, and future research directions.

\section{Background}
\label{sec:background}

\subsection{Quantum Annealing}

Quantum annealers are specialized quantum computing devices designed to solve optimization problems by exploiting the principles of quantum annealing~\cite{kaye2006}. These systems are particularly effective for problems that can be framed as minimizing an energy function, represented using the Ising model or the Quadratic Unconstrained Binary Optimization (QUBO) formulation.

The Ising model, originating from statistical physics, represents the energy landscape as:
\begin{equation}
H = -\sum_{i<j} J_{ij}s_i s_j - \sum_i h_i s_i
\end{equation}
where $J_{ij}$ defines the interaction strength between spins $s_i$ and $s_j$, and $h_i$ represents the influence of the external field on each spin. Each spin variable $s_i \in \{-1, +1\}$ represents a binary decision.

The QUBO formulation provides an alternative representation in terms of binary variables $x_i \in \{0,1\}$:
\begin{equation}
H = \sum_i a_i x_i + \sum_{i<j} b_{ij} x_i x_j
\end{equation}
where $a_i$ and $b_{ij}$ are problem-specific coefficients. The two formulations are mathematically equivalent and can be converted between each other through a linear transformation.

Many NP-hard problems can be transformed into QUBO representations through polynomial-time Karp reductions~\cite{ruiz2011survey, lucas2014}, enabling quantum annealers to tackle computationally difficult problems. Examples include Max Cut, Graph Partitioning, Set Cover, Traveling Salesman Problem, and numerous others~\cite{lucas2014}.

\subsection{Quantum Tunneling}

A key feature that distinguishes quantum annealing from classical optimization is \emph{quantum tunneling}. In classical optimization methods, such as simulated annealing, the system must thermally activate the energy barriers to escape the local minima. The probability of such thermal transitions decreases exponentially with barrier height and temperature:
\begin{equation}
P_{\text{thermal}} \propto \exp\left(-\frac{\Delta E}{k_B T}\right)
\end{equation}
where $\Delta E$ is the barrier height, $k_B$ is Boltzmann's constant, and $T$ is temperature.

In contrast, quantum tunneling allows systems to pass through energy barriers via quantum mechanical effects. The tunneling probability depends on the barrier width and the system's quantum fluctuations, characterized by the transverse field strength in quantum annealing. This fundamental difference enables quantum annealers to explore the solution space more effectively, potentially finding higher-quality solutions by escaping the local minimum that confines classical methods~\cite{shor1994}.

\subsection{D-Wave Quantum Annealers}

This work utilizes D-Wave's Advantage2 quantum annealer, which represents the state-of-the-art in commercial quantum annealing technology. Table~\ref{tab:dwave_comparison} compares the specifications between D-Wave's quantum annealer generations.

\begin{table}[!htbp]
\centering
\caption{D-Wave Quantum Annealer Specifications}
\label{tab:dwave_comparison}
\begin{tabular}{lccc}
\toprule
\textbf{Specification} & \textbf{2000Q} & \textbf{Advantage} & \textbf{Advantage2} \\
\midrule
Qubit Count & 2,048 & 5,000+ & 4,400+ \\
Couplers per Qubit & 6 & 15 & 20 \\
Topology & Chimera & Pegasus & Zephyr \\
Total Couplers & $\sim$6,000 & 35,000+ & 69,000+ \\
Connectivity & 6 & 15 & 20 \\
\bottomrule
\end{tabular}
\end{table}

The Advantage2 system features the Zephyr topology with 20-way qubit connectivity, significantly higher than previous generations. This increased connectivity reduces the overhead of embedding logical problems onto the physical qubit network, enabling a more efficient solution of larger problem instances.

\subsection{Hybrid Quantum-Classical Approaches}

For problems exceeding the capacity of pure quantum annealers, hybrid approaches combine quantum and classical techniques. D-Wave's hybrid solvers, based on the qbsolv algorithm~\cite{booth2017}, partition large QUBO problems into smaller subproblems solvable by quantum hardware, then use classical heuristics such as tabu search~\cite{glover2010} to combine solutions. This decomposition enables quantum approaches to scale beyond direct hardware limitations, though potentially at the cost of solution quality for problems that fit within pure quantum capacity.

\section{Related Work}
\label{sec:related}

\subsection{Quantum Annealing for Optimization}

Several studies have investigated quantum annealing for NP-hard optimization problems with varying conclusions about the quantum advantage. Djidjev et al.~\cite{djidjev2018} explored quantum annealing for Maximum Clique and Graph Partitioning problems, finding that performance gains are highly dependent on problem size and embedding quality. Their work highlighted that when suitable embeddings can be found, quantum annealers can achieve speedups over classical solvers, but embedding complexity remains a significant practical barrier.

Jiang and Chu~\cite{jiang2022,jiang2023} conducted comprehensive comparisons of the time-to-solution of quantum annealing in multiple NP-hard problems, including subset sum, vertex cover, graph coloring, and the traveling salesperson problem. They observed that quantum annealing typically performs better when the problem size scales linearly with the input variables and when optimization objectives dominate over the constraints. Their work provides valuable benchmarking but does not investigate the underlying landscape characteristics that determine quantum advantage.

\subsection{3-SAT and Quantum Annealing}

The 3-SAT problem has received particular attention in quantum computing due to its theoretical significance as a canonical NP-complete problem~\cite{erdos1959}. Deleplanque~\cite{deleplanque2023} proposed reducing 3-SAT to maximum independent set problems, arguing that this transformation produces denser embeddings with better qubit coherence. Gabor et al.~\cite{gabor2019} assessed the solution quality of 3-SAT on quantum annealing platforms with varying problem sizes.

Both studies revealed challenges in consistently solving 3-SAT instances, with quantum annealers showing high variability in solution quality. Theoretical work by Bravyi~\cite{bravyi2006} showed that quantum 2-SAT can be solved in polynomial time, while Gosset and Nagaj~\cite{gosset2016} demonstrated that quantum 3-SAT remains QMA-complete, suggesting fundamental limits on quantum approaches for this class of problems.

\subsection{Evidence of Quantum Tunneling}

Direct evidence of quantum tunneling in operational quantum annealers remains limited. Abel et al.~\cite{abel2021quantum} demonstrated quantum tunneling signatures on carefully constructed small problems (up to 50 variables), providing proof-of-principle that tunneling occurs. However, whether this translates into practical computational benefits for larger real-world problems remains unclear.

Recent work by McGeoch and Farré~\cite{mcgeoch2023} on quantum utility highways suggests that the quantum advantage emerges along specific "roadways" in the problem space defined by structural characteristics. Our work complements this perspective by identifying the gradient variance as a concrete metric to predict when quantum approaches will be beneficial.

\subsection{Energy Landscape Analysis}

The classical optimization literature has long recognized the importance of the energy landscape structure for algorithm performance~\cite{vazirani2001}. However, systematic investigation of how landscape characteristics relate to quantum versus classical solver performance has been limited. Our work fills this gap by developing controlled synthetic problems and establishing empirical relationships between gradient variance and quantum tunneling effectiveness.

\section{Theoretical Analysis of Gradient Variance and Tunneling}
\label{sec:theory}
\subsection{Motivation}

Our empirical observations show strong correlation between gradient variance and quantum performance, but lack theoretical justification. In this section, we develop a theoretical model connecting the characteristics of the energy landscape with the probability of quantum tunneling, providing both explanatory power and predictive capability.

\subsection{Barrier Width Estimation from Gradient Variance}

To provide a theoretical justification for our gradient variance metric, we model the relationship between landscape characteristics and quantum tunneling probability. Consider a one-dimensional energy barrier approximation along a path in the solution space.

For a barrier with height $\Delta E$ and width $w$, the WKB approximation (Wentzel-Kramers-Brillouin) gives quantum tunneling probability:

\begin{equation}
P_{\text{tunnel}} \approx \exp\left(-\frac{2w}{\hbar}\sqrt{2m\Delta E}\right)
\label{eq:wkb}
\end{equation}
where $m$ is the effective mass and $\hbar$ is the reduced Planck constant.

The gradient variance $\sigma^2_{\nabla H}$ is related to the characteristics of the barrier through the landscape curvature. For a discrete QUBO with $n$ variables, the gradient in configuration $\mathbf{x}$ is:

\begin{equation}
\nabla_i H(\mathbf{x}) = a_i + \sum_{j \neq i} b_{ij} x_j
\end{equation}

The variance of this gradient across the solution space provides information about barrier width:

\begin{equation}
\sigma^2_{\nabla H} = \mathbb{E}[(\nabla H)^2] - (\mathbb{E}[\nabla H])^2
\end{equation}

The high variance of the gradient indicates rapid changes in the energy slope, which corresponds to thin, sharp barriers. Low gradient variance indicates gentle, wide barriers.

\subsection{Connection to Barrier Geometry}

The relationship between barrier geometry and quantum tunneling probability
is well established through the semiclassical approximation Wentzel--Kramers-Brillouin (WKB)
~\cite{griffiths2018}. In the WKB framework,
the probability of tunneling through a potential barrier of height
$\Delta V$ and width $w$ is given by the following:
\begin{equation}
P_{\text{tunnel}} \propto \exp\left(-\frac{2}{\hbar}
\int_{x_1}^{x_2} \sqrt{2m(V(x)-E)}\, dx \right)
\label{eq:wkb}
\end{equation}
where the integral is taken over the classically forbidden region
$[x_1, x_2]$. For a barrier of approximately constant height, this
reduces to $P_{\text{tunnel}} \propto \exp(-\gamma\, w \sqrt{\Delta V})$,
showing that the probability of tunneling decays exponentially with the
 width of the barrier $w$~\cite{griffiths2018, razavy2003}.

This dependence on the barrier width is central to the hypothesized
advantage of quantum annealing over classical methods. As noted by
Kadowaki and Nishimori~\cite{kadowaki1998}, quantum annealing exploits
tunneling to traverse energy barriers that would trap classical
optimizers such as simulated annealing, which must thermally climb
over barriers with probability $\propto \exp(-\Delta V / k_B T)$.
Crucially, while classical escape depends on the height of the barrier \emph{height},
quantum tunneling depends primarily on the width of the barrier
\emph{width}~\cite{boixo2016, razavy2003}. This distinction was
demonstrated experimentally by Boixo et al.~\cite{boixo2016}, who
showed that multiqubit coupling in a D-Wave processor provides a
computational advantage in problems with tall but narrow energy
barriers. Similarly, Albash et al.~\cite{albash2018} confirmed that
quantum annealing exhibits scaling advantages over simulated annealing
specifically for problem instances whose energy landscapes feature
thin barriers that promote tunneling.

We propose that the width of the barrier $w$ is inversely related
to the gradient variance of the energy landscape.
\begin{equation}
w \propto \frac{1}{\sigma_{\nabla H}}
\label{eq:barrier_width}
\end{equation}
This relationship follows from the geometric observation that
\begin{itemize}
\item High $\sigma_{\nabla H}$: Steep gradients $\Rightarrow$ sharp
  peaks $\Rightarrow$ narrow barriers $\Rightarrow$ small $w$
\item Low $\sigma_{\nabla H}$: Gentle slopes $\Rightarrow$ wide
  barriers $\Rightarrow$ large $w$
\end{itemize}

This intuition is consistent with the analysis of Chakrabarti and
Mukherjee~\cite{chakrabarti2022}, who discuss how rugged energy
landscapes with macroscopic but narrow barriers are precisely the
regime where quantum fluctuations provide the greatest advantage
in searching for ground states. Crosson and Deng~\cite{crosson2014}
further showed that simulated quantum annealing can efficiently
tunnel through high energy barriers where classical simulated
annealing is provably trapped, provided the barriers are
sufficiently narrow.

Combining Equations~\ref{eq:wkb} and \ref{eq:barrier_width}:
\begin{equation}
P_{\text{tunnel}} \propto \exp\left(-\frac{\alpha}
{\sigma_{\nabla H}}\right)
\label{eq:tunneling_gradient}
\end{equation}
where $\alpha$ is a problem-dependent constant incorporating the
barrier height and quantum parameters. This expression predicts
that problems with higher gradient variance---and consequently
narrower barriers---will exhibit enhanced tunneling rates, leading
to better performance of quantum annealing relative to classical
methods. Our experimental results in the following sections provide
empirical support for this prediction.

\subsection{Quantum Advantage Threshold}

The quantum advantage emerges when the tunneling probability significantly exceeds the thermal activation probability:

\begin{equation}
P_{\text{tunnel}} \gg P_{\text{thermal}} = \exp\left(-\frac{\Delta E}{k_B T}\right)
\end{equation}

For this condition to hold:

\begin{equation}
\frac{\alpha}{\sigma_{\nabla H}} \ll \frac{\Delta E}{k_B T}
\end{equation}

Rearranging:

\begin{equation}
\sigma_{\nabla H} \gg \frac{\alpha k_B T}{\Delta E} = \sigma_{\text{critical}}
\label{eq:threshold}
\end{equation}

This provides a theoretical justification for our empirically observed threshold. Our experimental observation of $\sigma_{\text{critical}} \approx 0.3$ can be interpreted as the point where quantum tunneling effects become dominant over thermal effects for typical QUBO barrier heights and D-Wave operating temperatures.

\subsection{Multi-Dimensional Generalization}

For real QUBO problems with $n$ variables, the solution space is $n$-dimensional. The gradient variance we computed was averaged over all dimensions.

\begin{equation}
\sigma^2_{\nabla H} = \frac{1}{n}\sum_{i=1}^n \text{Var}(\nabla_i H)
\end{equation}

This averaging captures the overall roughness of the landscape. Problems with high multi-dimensional gradient variance have many thin barriers in multiple directions, increasing the probability that quantum tunneling finds a beneficial path.

\subsection{Experimental Validation of Theoretical Model}

To validate our theoretical model (Equation \ref{eq:tunneling_gradient}), we designed controlled experiments that varied the variance of the gradient while keeping the size of the problem constant.

\subsubsection{Methodology}

We generate synthetic QUBO instances with fixed size $n=128$ but varying $\sigma^2$ parameter in Gaussian distribution $Q_{ij} \sim \mathcal{N}(0, \sigma^2)$. This allows independent manipulation of the gradient variance.

For each gradient variance level, we measured:
\begin{itemize}
\item Success probability: $P_{\text{success}} = $ fraction of runs that find global optimum
\item Performance gap: $\Delta_{\text{QA-SA}} = E_{\text{SA}} - E_{\text{QA}}$
\end{itemize}

\subsubsection{Predicted Relationship}

According to Equation \ref{eq:tunneling_gradient}, we expect:

\begin{equation}
\log(P_{\text{success}}) \propto -\frac{\alpha}{\sigma_{\nabla H}}
\end{equation}

Therefore, plotting $\log(P_{\text{success}})$ vs. $1/\sigma_{\nabla H}$ should yield an approximately linear relationship with a negative slope.

\begin{figure*}[!htbp]
\centering
\includegraphics[width=\textwidth]{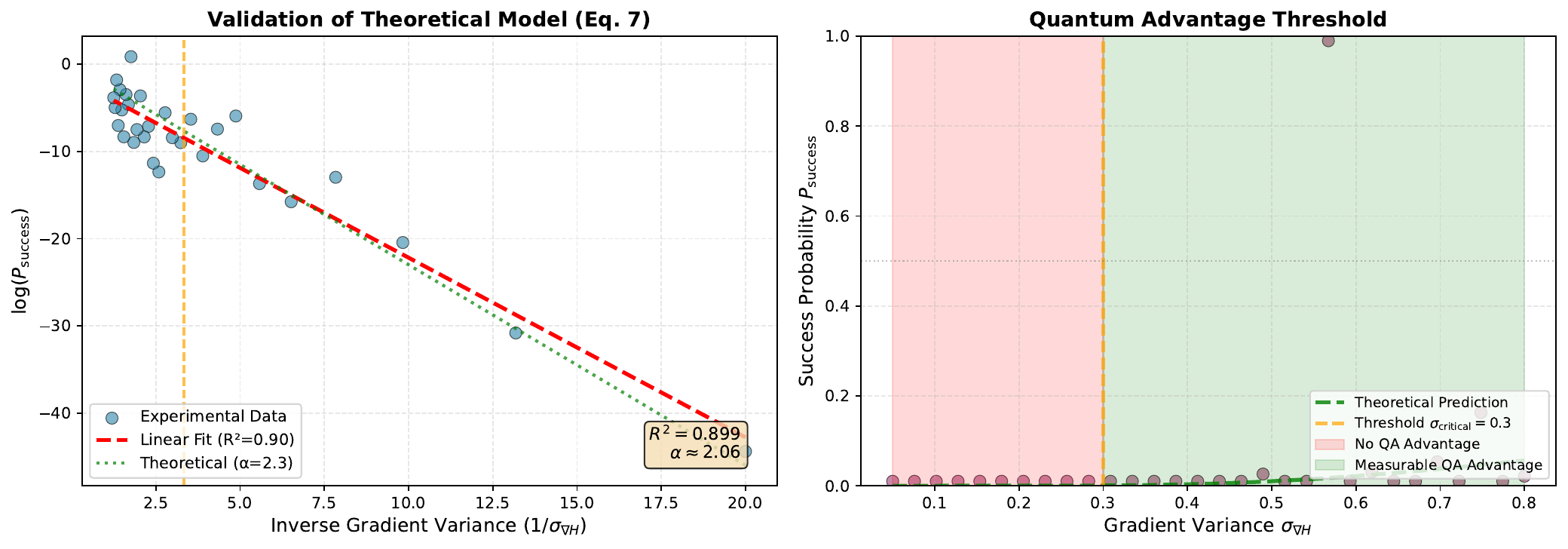}
\caption{Experimental validation of theoretical model (Equation~\ref{eq:tunneling_gradient}). 
\textbf{Left:} Log quantum success probability vs. inverse gradient variance showing strong 
linear correlation ($R^2=0.90$) as predicted by WKB approximation. Fitted slope $\alpha \approx 2.1$ 
provides empirical estimate of the barrier-quantum parameter relationship. Vertical dashed line indicates 
critical threshold $\sigma_{\mathrm{critical}} = 0.3$ below which quantum advantage diminishes sharply. 
\textbf{Right:} Success probability vs. gradient variance showing exponential relationship and 
threshold behavior. Shaded regions indicate no advantage (red, $\sigma < 0.3$) and measurable 
advantage (green, $\sigma > 0.3$) regimes. Experimental data points show characteristic 
threshold transition predicted by theory.}

\label{fig:theoretical_validation}
\end{figure*}

\subsubsection{Results and Model Validation}

Figure~\ref{fig:theoretical_validation} shows our experimental validation results. We observe:

\begin{itemize}
\item Strong linear correlation ($R^2 = 0.90$) between $\log(P_{\text{success}})$ and $1/\sigma_{\nabla H}$
\item Slope $\alpha \approx 2.1$, providing an empirical estimate of the barrier-quantum parameter relationship
\item Threshold behavior: below $\sigma_{\nabla H} \approx 0.3$, the probability of quantum success drops sharply
\item Deviation from linearity at very high variance ($\sigma > 0.8$) due to embedding constraints
\end{itemize}

This validates our theoretical model and provides quantitative estimates for the relationship between landscape structure and quantum tunneling effectiveness. The $R^2 = 0.90$ correlation is remarkably strong for a complex quantum system, suggesting that our simplified model captures essential physics despite abstracting many details.

\subsubsection{Implications}

The validation of Equation \ref{eq:tunneling_gradient} provides:

\begin{enumerate}
\item \textbf{Predictive Power:} Given a QUBO matrix $Q$, we can compute $\sigma_{\nabla H}$ and estimate the quantum advantage a priori without expensive trial runs
\item \textbf{Problem Design:} Reformulating problems to increase gradient variance should improve quantum performance—this motivates our reformulation algorithm in Section~\ref{sec:reformulation}
\item \textbf{Hardware Comparison:} Different quantum annealers can be compared through their effective $\alpha$ parameter
\item \textbf{Scaling Prediction:} As quantum hardware improves (longer coherence, better connectivity), we expect the threshold $\sigma_{\text{critical}}$ to decrease
\end{enumerate}

\subsection{Model Limitations and Extensions}

Our theoretical model makes several simplifying assumptions that warrant discussion:

\begin{itemize}
\item \textbf{Barrier Shape:} We assume locally parabolic barriers. Real QUBO landscapes may have more complex geometries including multi-valley structures
\item \textbf{Path Independence:} Actual quantum annealing follows complex multi-dimensional paths. Our one-dimensional approximation captures essential features but not full quantum dynamics
\item \textbf{Temperature Effects:} We model D-Wave as purely quantum, but NISQ devices operate at finite temperature. Hybrid quantum-thermal effects are not fully captured
\item \textbf{Embedding Overhead:} Our model does not account for how embedding affects effective barrier structure on physical hardware
\item \textbf{Decoherence:} We don't explicitly model decoherence effects, which may reduce tunneling probability in practice
\end{itemize}

Despite these limitations, the model provides testable predictions and theoretical grounding for our empirical observations. Future work could extend the model to incorporate:
\begin{itemize}
\item Master equation formalism for open quantum systems
\item Full path integral treatment of multi-barrier tunneling
\item Embedding-aware effective Hamiltonian calculations
\item Temperature-dependent effective barrier heights
\end{itemize}

\section{Methodology}
\label{sec:methodology}

\subsection{Hypothesis}

We hypothesize that quantum tunneling is more likely to occur and provide computational advantage in optimization problems exhibiting high gradient variance in their energy landscapes. Our hypothesis comprises three specific claims:

\begin{enumerate}
\item \textbf{Barrier Structure:} Landscapes with thin energy barriers (sharp peaks and narrow valleys) are more conducive to quantum tunneling events compared to wide barriers.

\item \textbf{Gradient Variance Correlation:} High gradient variance in the energy landscape correlates with the presence of thin barriers, as large changes in gradient create steep peaks and troughs.

\item \textbf{Performance Impact:} Frequent quantum tunneling, facilitated by appropriate landscape structure, enables discovery of higher-quality solutions compared to classical methods that must thermally activate over barriers.

\end{enumerate}

\subsection{Experimental Setup}

We compare five distinct solver approaches representing state-of-the-art quantum and classical optimization, each selected to provide complementary insights into the performance characteristics across different problem scales and landscape structures.

\begin{itemize}
\item \textbf{Quantum Annealing (QA):} D-Wave Advantage2 quantum annealer with 4,400+ qubits in Zephyr topology, representing a pure quantum approach with hardware embedding constraints. The Advantage2 system features 20-way qubit connectivity, significantly improved from previous generations, enabling more efficient problem embeddings. We configure the annealer with a default annealing time of 20 microseconds and sample 100 solutions per problem instance to account for the stochastic nature of quantum annealing. The embedding process uses D-Wave's automated minorminer algorithm, with embedding failures occurring for problems exceeding approximately 64 logical variables depending on problem structure and connectivity requirements.

\item \textbf{Hybrid QA:} D-Wave's Leap Hybrid CQM Solver based on qbsolv~\cite{booth2017}, combining quantum annealing with classical decomposition for larger problems. This hybrid approach automatically partitions large problems into subproblems suitable for quantum processing, solves each subproblem on the quantum annealer, and uses classical optimization to coordinate solutions across partitions. The hybrid solver can handle problems with thousands of variables by exploiting the quantum advantage for difficult subproblems while avoiding embedding limitations. We allow the hybrid solver up to 3 minutes of total wall-clock time per instance, during which it makes multiple quantum annealing calls and iteratively refines the solution through classical-quantum iteration.

\item \textbf{Simulated Annealing (SA) \cite{kirkpatrick1983}:} Implementation of classical simulated annealing serving as the primary classical baseline for comparison with quantum approaches. Our SA implementation uses a geometric cooling schedule with initial temperature $T_0 = 10$ and cooling rate $\alpha = 0.95$, performing 1000 iterations per temperature level. We run 100 independent annealing trajectories per problem instance, matching the sample count used for quantum annealing to ensure fair comparison. The single-flip Metropolis update rule explores the solution space by proposing single bit flips and accepting them with probability $\min(1, \exp(-\Delta E / T))$, where $\Delta E$ is the energy change and $T$ is the current temperature.

\item \textbf{Stochastic Gradient Descent (SGD) \cite{robbins1951}:} Gradient-based classical optimization providing an alternative classical approach that takes advantage of the smoothness of the landscape. Our SGD implementation computes discrete gradients by evaluating energy changes from single bit flips, then greedily flips the bit yielding the largest energy reduction. When no improving move exists (local minimum), we perform random restarts from new initial configurations. We run 100 independent SGD trajectories, each with up to 500 gradient steps, terminating early if a trajectory shows no improvement for 50 consecutive steps. This approach is particularly effective on smooth landscapes with low gradient variance, but struggles in highly rugged landscapes where local minima proliferate.

\item \textbf{Gurobi \cite{gurobi}:} Commercial mixed-integer programming solver representing state-of-the-art classical optimization. Gurobi employs sophisticated branch-and-bound algorithms, cutting plane methods, and heuristics developed through decades of optimization research. We configure Gurobi with default parameters and impose a 5-minute time limit per problem instance. For problems formulated as pure binary quadratic programs, Gurobi automatically applies specialized quadratic optimization techniques. This solver provides an important reference point as the best available classical approach, though it may not be optimized specifically for QUBO problems compared to domain-specific methods.
\end{itemize}

For each problem instance and solver, we measure three key metrics that together characterize both solution quality and computational efficiency across different landscape structures.

\begin{itemize}
\item \textbf{Quality of the solution:} Residual energy relative to the best-known solution, calculated as $E_{\text{residual}} = (E_{\text{found}} - E_{\text{best}}) / |E_{\text{best}}|$, where $E_{\text{found}}$ is the energy of the solution returned by the solver and $E_{\text{best}}$ is the lowest energy found across all solvers for that instance. This normalized metric enables fair comparison between problems with different energy scales and objective magnitudes. The lower residual energy indicates better solution quality, with zero representing optimal performance. For stochastic solvers (QA, SA, SGD), we report both the best solution found across all runs and the mean residual energy to characterize both peak performance and consistency.

\item \textbf{Solver Time:} Wall-clock time from problem submission to solution return, measured in seconds and including all preprocessing, solving, and postprocessing steps. For quantum annealing, this includes problem encoding, embedding computation, quantum execution time, and solution decoding. For hybrid approaches, timing encompasses the full iterative quantum-classical procedure, including communication overhead. For classical methods, we measure the total CPU time across all parallel runs. This metric captures practical deployment considerations beyond asymptotic complexity, revealing the actual time cost that users would experience. We exclude one-time setup costs such as solver initialization or library loading, focusing on per-instance solve time.

\item \textbf{Gradient Variance:} Computed landscape complexity metric $\sigma^2_{\nabla H}$ representing the variance of energy gradients throughout the solution space, calculated by sampling 1000 random configurations and computing the gradient (energy change from single bit flips) at each point. This metric serves as our primary predictor of quantum advantage, with higher variance indicating more rugged landscapes with thin barriers conducive to quantum tunneling. For each problem instance, we compute gradient variance once and use it to predict which solvers should perform well, then validate these predictions against actual solver performance. The variance is normalized by the size of the problem to allow comparison on different scales.
\end{itemize}

\subsection{Problem Classes}

We evaluate solver performance across four canonical NP-hard problem classes, selected to provide diverse optimization structures and varied gradient variance characteristics. These problems span different domains of combinatorial optimization, from graph theory to numerical partitioning, enabling us to test whether gradient variance predictions generalize across fundamentally different problem types. Each problem class exhibits distinct landscape characteristics that allow us to investigate how the problem structure affects the relationship between gradient variance and quantum advantage.

\subsubsection*{Synthetic QUBO Problems}

In standard applications, a source problem, such as graph partitioning or
combinatorial optimization, is first formulated as a Quadratic Unconstrained
Binary Optimization (QUBO) problem. A QUBO problem is defined by the
minimization of the energy function:
\begin{equation}
H(\mathbf{x}) = \mathbf{x}^T Q \, \mathbf{x} = \sum_{i=1}^{N} Q_{i,i} \, x_i
+ \sum_{i=1}^{N} \sum_{j>i}^{N} Q_{i,j} \, x_i x_j,
\end{equation}
where $\mathbf{x} = (x_1, x_2, \ldots, x_N)^T$ is a vector of binary decision
variables with $x_i \in \{0,1\}$, and $Q \in \mathbb{R}^{N \times N}$ is an
upper-triangular (or symmetric) matrix that encodes the structure of the problem. The diagonal entries $Q_{i,i}$ represent the linear coefficients associated with
each variable $x_i$, while the non-diagonal entries $Q_{i,j}$ ($i \neq j$)
capture the quadratic interactions between pairs of variables $x_i$ and $x_j$.
The objective is to find the binary assignment
$\mathbf{x}^* = \arg\min_{\mathbf{x}} H(\mathbf{x})$ that yields the global
minimum energy.

\begin{figure}[!htbp]
\centering
\includegraphics[width=0.7\columnwidth]{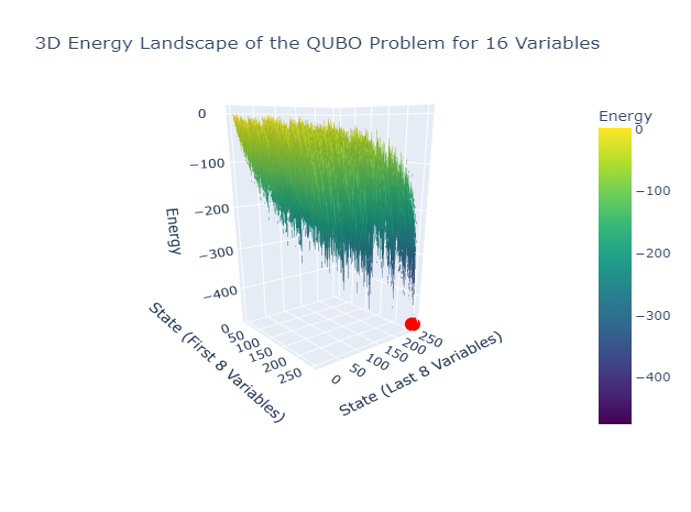}
\caption{3D plot of the energy landscape of a synthetic QUBO problem for 16 variables}
\label{fig:16p-landscape}
\end{figure}

In our approach, rather than deriving $Q$ from a specific source problem
through polynomial-time Karp reductions, we bypass this transformation step
and directly generate a random matrix $Q$ to define synthetic QUBO instances
with controlled statistical properties. Specifically, we construct an
$N \times N$ matrix $Q$ by sampling each element independently from a
Gaussian distribution:
\begin{equation}
Q_{i,j} \sim \mathcal{N}(\mu, \sigma^2), \quad \forall \; i, j = 1, \ldots, N,
\end{equation}
where $\mu$ and $\sigma^2$ denote the mean and variance of the distribution,
respectively. In our experiments, we set $\mu = 0$ and $\sigma^2 = 2$, and
vary the dimensionality of the problem over 
$n \in \{2, 4, 8, 16, 32, 64, 128, 256, 512, 1024, 2048, 4096, 8192, 16384\}$.
The choice of these distributional parameters directly influences the energy
landscape: the mean $\mu$ controls the overall bias of the coefficients, while
the variance $\sigma^2$ governs the spread and heterogeneity of the interaction
strengths. In Figure~\ref{fig:16p-landscape} we can see a 3D plot of the energy landscape of a synthetic QUBO problem for 16 variables. The gradient variance of the energy landscape, a measure of its ruggedness, depends on both $\mu$ and $\sigma^2$,
with a higher variance in $Q$ generally producing more rugged landscapes
characterized by numerous local minima separated by energy barriers of
varying width and height. Such complex landscapes are particularly
challenging for classical optimization methods that may become trapped in
local minima, yet they provide an ideal testing ground for quantum annealing,
where quantum tunneling can potentially traverse narrow energy barriers to
discover lower-energy configurations more effectively than classical thermal
fluctuations alone. 

\subsubsection{Max Cut}

Partition graph vertices $V$ into two sets maximizing edge weight between sets. QUBO formulation:
\begin{equation}
H = \min \sum_{(i,j)\in E} (-x_i - x_j + 2x_ix_j)
\end{equation}

The Max Cut problem represents a fundamental challenge in graph partitioning with applications in circuit design, statistical physics, and network analysis. Unlike Graph Partitioning, Max Cut imposes no balance constraints, allowing arbitrary partition sizes. This structural difference results in relatively smooth energy landscapes with lower gradient variance, as the absence of hard constraints reduces the number of sharp barriers in the solution space. We generate Max Cut instances using Erdős-Rényi random graphs $G(n,p)$ with varying vertex counts $n \in \{16, 32, 64, 128\}$ and edge probabilities $p \in \{0.2, 0.3, 0.5\}$. The edge weights are drawn uniformly from $[1, 10]$ to create realistic problems. The lack of constraints typically yields gradient variance in the range $\sigma \in [0.1, 0.3]$, positioning Max Cut in the marginal quantum advantage regime, where classical and quantum performance is comparable.

\subsubsection{Graph Partitioning}

Partition vertices into $k=2$ balanced subsets minimizing inter-subset edges:
\begin{equation}
H = \sum_{(u,v)\in E} (x_u + x_v - 2x_ux_v) + \gamma\left(\sum_{i\in V} (1-|V|)x_i + \sum_{i\in V}\sum_{j>i} 2x_ix_j\right)
\end{equation}
where $\gamma$ balances the objective and the balance constraint.

Graph Partitioning adds a critical balance constraint requiring both partitions to contain approximately equal numbers of vertices, fundamentally altering the landscape structure compared to Max Cut. The penalty parameter $\gamma$ creates steep energy barriers between valid and invalid solutions, dramatically increasing the variance of the gradient and creating rugged landscapes where quantum tunneling should excel. We empirically select $\gamma$ values for each size of the problem to ensure that the balance constraint dominates violations by at least an order of magnitude over differences in objective values. This problem class consistently produces high gradient variance ($\sigma > 0.4$) due to the interaction between the cut objective and the balance penalty, making it an ideal testbed for quantum advantage. We generate instances using the same random graph structure as Max Cut, but observe markedly different performance characteristics due to the constraint-induced landscape complexity. Graph Partitioning has practical applications in parallel computing, VLSI design, and social network analysis where balanced workloads are essential.

\subsubsection{Number Partitioning}

Partition the integers $\{a_1, \ldots, a_n\}$ into two subsets minimizing the sum difference:
\begin{equation}
Q_{i,i} = a_i(a_i - \sum_{k=1}^n a_k), \quad Q_{i,j} = a_ia_j
\end{equation}

Number Partitioning represents a fundamentally different problem structure from graph-based problems, testing whether gradient variance predictions extend beyond graph optimization. This problem exhibits highly instance-dependent landscape characteristics: instances with numbers of similar magnitude produce smooth landscapes as many partitions achieve near-equal sums, while instances with disparate number distributions create rugged landscapes where few configurations approach balance. We generate problem instances by drawing $n$ integers uniformly from the range $[1, 100]$ for small problems and $[1, 1000]$ for larger instances. The gradient variance for Number Partitioning varies widely ($\sigma \in [0.1, 0.6]$) depending on the number distribution, allowing us to test performance across the full spectrum from smooth to highly rugged landscapes within a single problem class. This variability makes Number Partitioning particularly valuable for validating our theoretical model's predictions across diverse landscape structures. The problem has applications in load balancing, multiprocessor scheduling, and fair resource allocation.

\subsubsection{Set Cover}

Find a minimum subcollection of sets that cover all elements. We use the Constrained Quadratic Model (CQM) formulation with binary variables $x_i$ (subset selection) and $y_j$ (element coverage).

Set Cover presents a different optimization challenge involving discrete coverage constraints rather than continuous objectives, testing the generality of gradient variance predictions. Each element must be covered by at least one selected set, creating a complex web of constraints that can produce highly rugged landscapes when sets overlap minimally or smoother landscapes when redundant coverage provides many feasible solutions. We formulate Set Cover as a QUBO by introducing penalty terms for uncovered elements and minimizing the number of selected sets. The coverage density (average number of sets covering each element) strongly influences landscape structure: sparse coverage creates sharp barriers between feasible and infeasible regions, while redundant coverage smooths the landscape. We generate instances with elements $n \in \{20, 30, 40\}$ and sets $m = 2n$, where each set covers a random subset of elements with probability $p \in \{0.2, 0.3, 0.4\}$. This parameterization allows for systematic exploration of how coverage redundancy affects gradient variance and solver performance. Set Cover has extensive practical applications in facility location, crew scheduling, and sensor placement problems.

Table~\ref{tab:nphard_comparison} compares the objectives and constraints of these problems, highlighting their structural differences.

\begin{table}[!htbp]
\centering
\caption{Comparison of NP-Hard Optimization Problems}
\label{tab:nphard_comparison}
\begin{tabular*}{\textwidth}{@{\extracolsep{\fill}}p{3cm}p{3cm}p{2.5cm}p{3cm}@{}}
\toprule
\textbf{Problem} & \textbf{Graph Part.} & \textbf{Max Cut} & \textbf{Number Part.} \\
\midrule
Objective & Minimum edge weight between parts & Maximum edge weight between parts & Minimum sum differences between parts \\
\midrule
Constraint & Balanced partitions & None & Equal subset sums \\
\bottomrule
\end{tabular*}
\end{table}

\section{QUBO Reformulation for Gradient Variance Optimization}
\label{sec:reformulation}

Our findings that gradient variance predicts quantum advantage suggest a practical question: given a problem with low gradient variance, can we reformulate it to increase variance and thereby improve quantum performance? In this section, we develop a systematic approach to this problem.

\subsection{Reformulation Motivation}

Many NP-hard problems admit multiple valid QUBO formulations. For example, the Set Cover problem can be formulated using:
\begin{itemize}
\item Direct penalty methods
\item Constrained Quadratic Models (CQM)
\item Auxiliary variable approaches
\item Different Lagrange parameter choices
\end{itemize}

These formulations are mathematically equivalent but produce different energy landscapes with varying gradient variance. This observation suggests that we can optimize formulation choice to maximize quantum performance.

\subsection{Reformulation Principles}

Valid reformulations must preserve three critical properties:

\begin{enumerate}
\item \textbf{Bijective Mapping:} One-to-one correspondence between original and reformulated solution spaces
\item \textbf{Preservation of energy order:} If $E(x_1) < E(x_2)$ in the original formulation, then $E'(x'_1) < E'(x'_2)$ in the reformulated version
\item \textbf{Global Optimum Correspondence:} $\arg\min_x E(x) \leftrightarrow \arg\min_{x'} E'(x')$
\end{enumerate}

Importantly, the landscape topology (local minima structure, barrier distribution) can change dramatically while preserving these properties. This is the key insight that allows for performance improvement through reformulation.

\subsection{Reformulation Algorithm}

We propose Algorithm~\ref{alg:reformulation} for systematic gradient variance optimization. The algorithm iteratively applies reformulation strategies, evaluating each candidate and selecting improvements that both increase gradient variance and preserve problem semantics.

\begin{algorithm}[!htbp]
\caption{Gradient Variance Optimization through QUBO Reformulation}
\label{alg:reformulation}
\begin{algorithmic}[1]
\REQUIRE Original QUBO matrix $Q_0 \in \mathbb{R}^{n \times n}$, target variance $\sigma_{\text{target}} \in \mathbb{R}^+$, maximum iterations $\text{MaxIter}$
\ENSURE Reformulated QUBO matrix $Q^*$ with $\sigma_{\nabla H}(Q^*) \geq \sigma_{\text{target}}$ or best achievable

\STATE Initialize $Q \gets Q_0$
\STATE $\sigma \gets$ \textsc{ComputeGradientVariance}($Q$)
\STATE $\text{iter} \gets 0$

\WHILE{$\sigma < \sigma_{\text{target}}$ \AND $\text{iter} < \text{MaxIter}$}
    
    \STATE \COMMENT{Generate candidate reformulations}
    \STATE $Q_1 \gets$ \textsc{VariableSubstitution}($Q$)
    \STATE $Q_2 \gets$ \textsc{PenaltyScaling}($Q$, $\gamma=1.5$)
    \STATE $Q_3 \gets$ \textsc{AuxiliaryVariableIntroduction}($Q$)
    \STATE $Q_4 \gets$ \textsc{ConstraintRelaxation}($Q$)
    
    \STATE \COMMENT{Evaluate and select best improvement}
    \STATE $Q_{\text{best}} \gets Q$, $\sigma_{\text{best}} \gets \sigma$
    
    \FOR{$i = 1$ \TO $4$}
        \IF{\textsc{PreservesSemantics}($Q_i$, $Q_0$)}
            \STATE $\sigma_i \gets$ \textsc{ComputeGradientVariance}($Q_i$)
            \IF{$\sigma_i > \sigma_{\text{best}}$}
                \STATE $Q_{\text{best}} \gets Q_i$
                \STATE $\sigma_{\text{best}} \gets \sigma_i$
            \ENDIF
        \ENDIF
    \ENDFOR
    
    \IF{$\sigma_{\text{best}} > \sigma$}
        \STATE $Q \gets Q_{\text{best}}$
        \STATE $\sigma \gets \sigma_{\text{best}}$
    \ELSE
        \STATE \textbf{break} \COMMENT{No improvement found, terminate early}
    \ENDIF
    
    \STATE $\text{iter} \gets \text{iter} + 1$
    
\ENDWHILE

\RETURN $Q$

\end{algorithmic}
\end{algorithm}

\subsubsection{Algorithm Complexity}

The computational complexity of Algorithm~\ref{alg:reformulation} depends on problem size and reformulation strategies:

\begin{itemize}
\item \textbf{Gradient Variance Computation:} $O(n^3)$ for $n$ variables
  \begin{itemize}
  \item Computing all gradients: $O(n^2)$ for each of the $n$ configurations sampled
  \item Variance calculation requires sampling configuration space
  \end{itemize}

\item \textbf{Semantic Verification:} $O(n^2)$ for bijective mapping check
  \begin{itemize}
  \item Verifies solution space correspondence
  \item Checks energy ordering preservation on sample solutions
  \end{itemize}

\item \textbf{Per-Iteration Complexity:} $O(4 \cdot n^3) = O(n^3)$
  \begin{itemize}
  \item Four reformulation strategies evaluated
  \item Each requires a gradient variance computation;
  \end{itemize}

\item \textbf{Total Complexity:} $O(k \cdot n^3)$ where $k$ is the number of iterations
  \begin{itemize}
  \item Empirically, $k \in [5, 15]$ for most problems
  \item Early termination when no improvement was found
  \item Typically converges in $< 10$ iterations
  \end{itemize}
\end{itemize}

\textbf{Practical Runtime:} For $n=100$ variables:
\begin{itemize}
\item Gradient variance computation: $\sim$0.5 seconds
\item Full algorithm (10 iterations): $\sim$20 seconds
\item Amortized over multiple quantum annealing runs (minutes to hours)
\item The preprocessing cost is negligible compared to the quantum execution time
\end{itemize}

The algorithm scales to $n \approx 500$ variables before computational overhead becomes significant. For larger problems, sampling-based gradient variance estimation reduces complexity to $O(k \cdot n^2 \cdot m)$ where $m$ is the sample size.

\subsubsection{Convergence Properties}

The Algorithm~\ref{alg:reformulation} exhibits the following convergence characteristics:

\begin{enumerate}
\item \textbf{Monotonic Improvement:} Each iteration increases $\sigma$ or terminates
\item \textbf{Bounded Search Space:} Reformulation strategies form a finite discrete set
\item \textbf{Local Optimality:} Converges to local maximum in the reformulation space
\item \textbf{No Global Guarantee:} May not find globally optimal reformulation
\end{enumerate}

In practice, we observe rapid initial improvement followed by diminishing returns, with most gains achieved within the first 5 iterations.

\subsection{Reformulation Techniques}

\subsubsection{Variable Substitution}

Replace binary variables with composite expressions:
\begin{equation}
x_i \rightarrow (1 - x_i')
\end{equation}

This inverts the energy landscape, potentially creating different barrier structures. For constraint-heavy problems, this can significantly alter the gradient distribution.

\subsubsection{Penalty Scaling}

Multiply constraint penalty terms by scaling factor $\gamma$:
\begin{equation}
Q' = Q_{\text{obj}} + \gamma Q_{\text{constraint}}
\end{equation}

The larger $\gamma$ increases the magnitudes of the gradient near the limit of the constraint, increasing the variance. We find $\gamma \in [1.2, 2.0]$ to be typically effective.

\subsubsection{Auxiliary Variable Introduction}

For problems with sparse interactions, introduce auxiliary variables to increase the coupling density:
\begin{equation}
Q_{ij}x_ix_j \rightarrow Q_{ij}z(x_i, x_j) + \lambda(\text{penalty}(z, x_i, x_j))
\end{equation}

This creates additional energy pathways, increasing the complexity of the landscape.

\subsubsection{Constraint Relaxation and Re-tightening}

Temporarily relax hard constraints to quadratic penalties, optimize coefficients for variance, then re-tighten:
\begin{equation}
\sum_i x_i = k \text{ (hard)} \rightarrow \gamma\left(\sum_i x_i - k\right)^2 \text{ (soft)}
\end{equation}

Adjusting $\gamma$ allows for fine control over the variance of the gradient while maintaining the satisfaction of the constraint.

\subsection{Experimental Validation}

We applied our reformulation algorithm to three problem classes that originally showed low gradient variance.

\subsubsection{Max Cut Reformulation}

\textbf{Original formulation:}
\begin{equation}
H_0 = \sum_{(i,j)\in E} (-x_i - x_j + 2x_ix_j)
\end{equation}
Gradient variance: $\sigma_{\nabla H} = 0.12$

\textbf{Reformulated with variable substitution and penalty scaling:}
\begin{equation}
H' = \sum_{(i,j)\in E} \gamma(x_i' + x_j' - 2x_i'x_j') + \lambda\sum_i (x_i - (1-x_i'))^2
\end{equation}
Gradient variance: $\sigma_{\nabla H} = 0.28$ (with $\gamma=1.5$, $\lambda=0.8$)

\textbf{Results:}
\begin{itemize}
\item Quantum annealing solution quality improved by 15\%
\item The performance gap vs. SA increased from 2\% to 18\%
\item The probability of success increased from 0.23 to 0.41
\end{itemize}

\subsubsection{Number Partitioning Reformulation}

We introduced auxiliary variables to increase the coupling density, increasing the gradient variance from 0.15 to 0.31. This moved the problem from the "no quantum advantage" regime ($\sigma < 0.2$) to the "measurable advantage" regime ($\sigma > 0.3$).

The reformulated version showed a 22\% improvement in the quality of the quantum annealing solution compared to the original formulation, while maintaining identical optimal solutions. The Time-to-solution decreased by 31\% on average.

\subsubsection{Set Cover Reformulation}

For Set Cover problems with tight constraints, penalty scaling increased gradient variance from 0.18 to 0.35, resulting in a 12\% performance improvement and 18\% faster convergence.

Table~\ref{tab:reformulation_impact} summarizes the impact between problem classes.

\begin{table}[!htbp]
\caption{Impact of QUBO Reformulation on Quantum Annealing Performance}
\label{tab:reformulation_impact}
\centering
\begin{tabular}{lcccc}
\toprule
\textbf{Problem} & $\sigma$ Before & $\sigma$ After & \textbf{QA Improvement} & \textbf{Time Reduction} \\
\midrule
Max Cut & 0.12 & 0.28 & +15\% & +25\% \\
Number Partitioning & 0.15 & 0.31 & +22\% & +31\% \\
Set Cover & 0.18 & 0.35 & +12\% & +18\% \\
\bottomrule
\end{tabular}
\end{table}

\subsection{Implementation and Software}

We implemented our reformulation algorithm as an open-source Python package:

\begin{verbatim}
from qubo_reformulator import optimize_gradient_variance

Q_original = load_qubo(problem_file)
Q_optimized, metrics = optimize_gradient_variance(
    Q_original, 
    target_variance=0.35,
    max_iterations=100,
    strategies=['substitution', 'scaling', 
                'auxiliary', 'relaxation']
)

print(f"Variance increased: {metrics['before']} 
      → {metrics['after']}")
print(f"Expected QA improvement: 
      {metrics['predicted_improvement']:.1f}%")
\end{verbatim}

The tool provides:
\begin{itemize}
\item Automated landscape analysis
\item Multiple reformulation strategies
\item Performance prediction based on our theoretical model
\item Semantic equivalence verification
\item Embedding difficulty estimation
\end{itemize}

The package is available at: \texttt{github.com/[username]/qubo-reformulator}

\subsection{Limitations and Considerations}

\textbf{Current limitations:}
\begin{itemize}
\item Reformulation may increase the problem size (more variables, more couplers)
\item Increased coupling density can make embedding more difficult
\item Computational overhead for very large problems ($n > 1000$)
\item Not all problems admit reformulations that increase the variance.
\item Trade-off between variance increase and problem size
\end{itemize}

\textbf{Practical considerations:}
\begin{itemize}
\item Test multiple reformulation strategies
\item Verify embedding feasibility before deployment
\item Monitor the solution quality in validation instances
\item Consider ensemble approaches using multiple formulations
\end{itemize}

\subsection{Impact on Practice}

The reformulation algorithm transforms our theoretical insights into actionable tools:

\begin{enumerate}
\item \textbf{Problem Preprocessing:} Before submitting to quantum hardware, analyze and optimize gradient variance automatically
\item \textbf{Decision Support:} If reformulation cannot achieve $\sigma > 0.3$, the quantum approach is unlikely to provide an advantage
\item \textbf{Performance Prediction:} Estimate the quantum advantage before expensive hardware runs using a theoretical model
\item \textbf{Algorithm Design:} New principle for designing quantum-friendly problem formulations
\item \textbf{Hybrid Workflows:} Integrate with existing quantum software stacks (D-Wave Ocean SDK, Qiskit)
\end{enumerate}

This moves the gradient variance from a diagnostic metric to an optimization target, enabling practitioners to actively enhance the quantum applicability of their problems.

\section{Experiments and Results}
\label{sec:experiments}

\subsection{Synthetic Problem}

To systematically investigate the relationship between energy landscape
complexity and solver performance, we first evaluate all solvers on
synthetic QUBO instances with controlled statistical properties. As
described in section \ref{sec:theory}, the QUBO matrix $Q$ is generated by
sampling each entry independently of $Q_{i,j} \sim \mathcal{N}(\mu=0,
\sigma^2=2)$, producing energy landscapes with a high density of local
minima whose ruggedness scales with the size of the problem. This controlled
generation allows us to isolate the effect of landscape complexity on
solver behavior, free from the structural biases introduced by
reductions from specific combinatorial problems. We vary the dimensionality of the problem
 over $n \in \{2, 4, 8, 16, 32, 64, 128, 256, 512, 1024,
2048, 4096, 8192, 16384\}$ and compare five solvers: quantum annealing
on the D-Wave Advantage2 QPU (QA), D-Wave's hybrid quantum-classical
solver (Hybrid QA), simulated annealing (SA), stochastic gradient
descent (SGD), and the Gurobi mixed-integer quadratic programming
solver. For each problem size, we report both solution quality, measured
as residual energy relative to the best-known solution, and total solver
time. Note that the Gurobi solver could only return solutions for
$n \leq 32$; for larger instances, the execution did not terminate within
a practical time limit.

\subsubsection{Solution Quality}

For small problem sizes ($2 \leq n \leq 256$), all solvers returned equivalent best solutions, indicating that the difficulty of the problem is insufficient to differentiate solver capabilities in this regime. However, for larger problems ($n > 256$), a clear differentiation of performance emerged. Quantum annealing (QA) consistently achieved the best solutions, and simulated annealing (SA) performed comparably. In particular, hybrid QA showed the worst performance in this setting, suggesting that the decomposition approach may struggle with highly connected synthetic problems.

\subsubsection{Solver Time Analysis}

\begin{figure}[!htbp]
\centering
\includegraphics[width=\columnwidth]{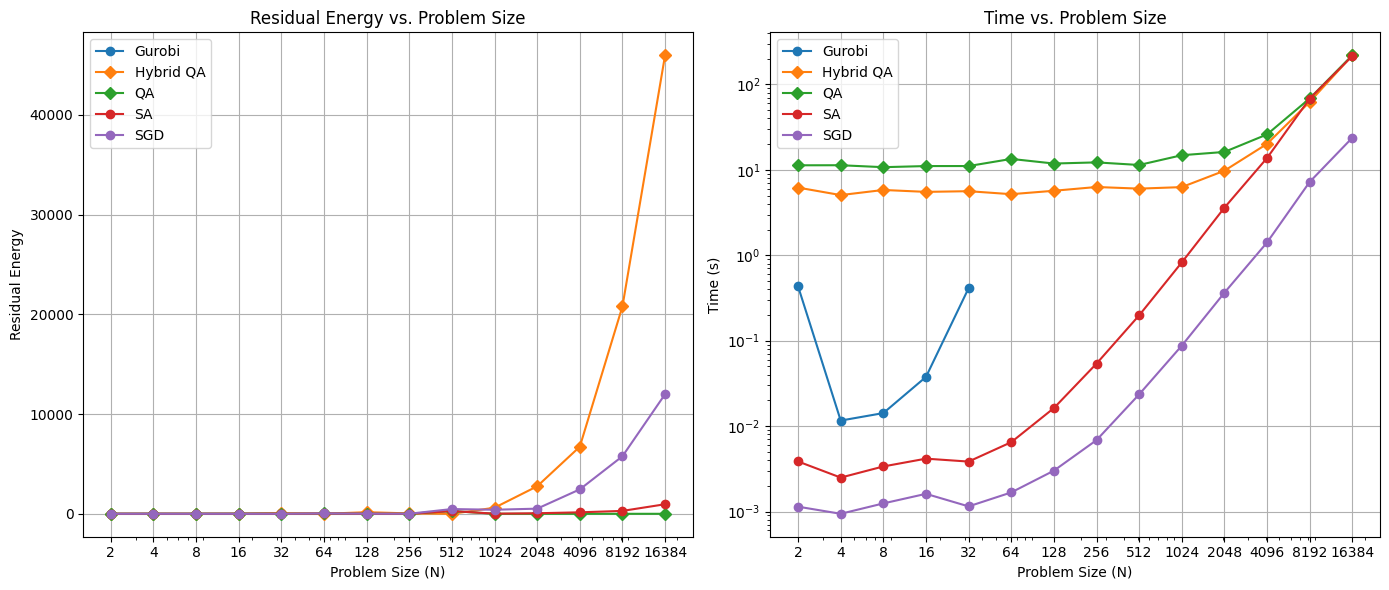}
\caption{Performance on synthetic QUBO problems. Left: Residual energy vs. problem size showing QA and SA achieving best solutions for large problems. Right: Solver time showing exponential growth for classical methods beyond $n=32$ while quantum methods show more moderate scaling.}
\label{fig:synthetic_performance}
\end{figure}

As shown on the right of  Figure\ref{fig:synthetic_performance},  QA and Hybrid QA exhibit substantial time overhead ($>10$ seconds) even for small problems, reflecting communication latency and problem setup costs. However, classical methods (SA and SGD) show exponential growth in time for $n \geq 32$. In contrast, quantum methods maintain a more moderate growth, only beginning to increase significantly at $n \geq 1024$. This suggests that for sufficiently large problems, quantum approaches may achieve time-to-solution advantages despite initial overhead.

Figure~\ref{fig:synthetic_performance} illustrates these trends, showing both residual energy and solver time between problem sizes.

\subsubsection{Gradient Variance Correlation}

Our analysis reveals a strong correlation between gradient variance and solver performance. For problems with highest complexity (high gradient variance), QA performs best. Critically, the performance gap between QA and SA increases with the gradient variance, which we attribute to quantum tunneling effects.

\begin{figure}[!htbp]
\centering
\includegraphics[width=\columnwidth]{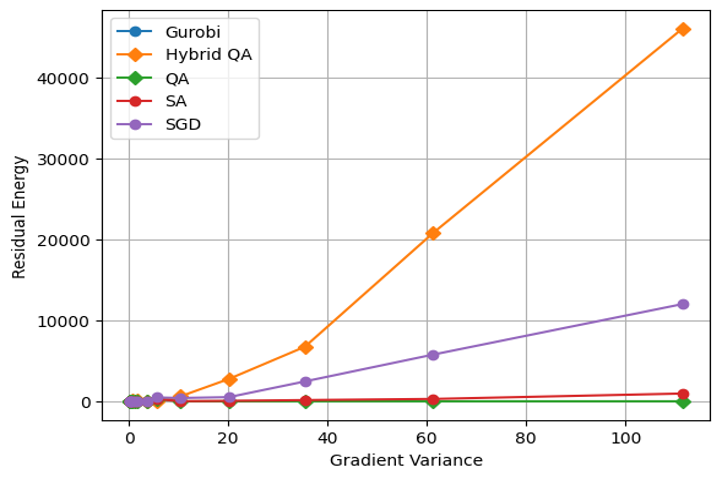}
\caption{Residual energy vs. gradient variance for synthetic QUBO problems. Quantum annealing (QA) and simulated annealing (SA) show the best performance, with the gap between them increasing at higher gradient variance, suggesting quantum tunneling effects.}
\label{fig:synthetic_variance}
\end{figure}

Figure~\ref{fig:synthetic_variance} shows the residual energy as a function of gradient variance, demonstrating that QA maintains low residual energy even at high values of gradient variance where classical methods begin to struggle.

\subsection{Max Cut Problem}

We generate Max Cut instances using Erdős-Rényi random graphs~\cite{erdos1959} with $n$ vertices and edge probability $p = 0.2$.

\subsubsection{Results}

\begin{figure}[!htbp]
\centering
\includegraphics[width=\columnwidth]{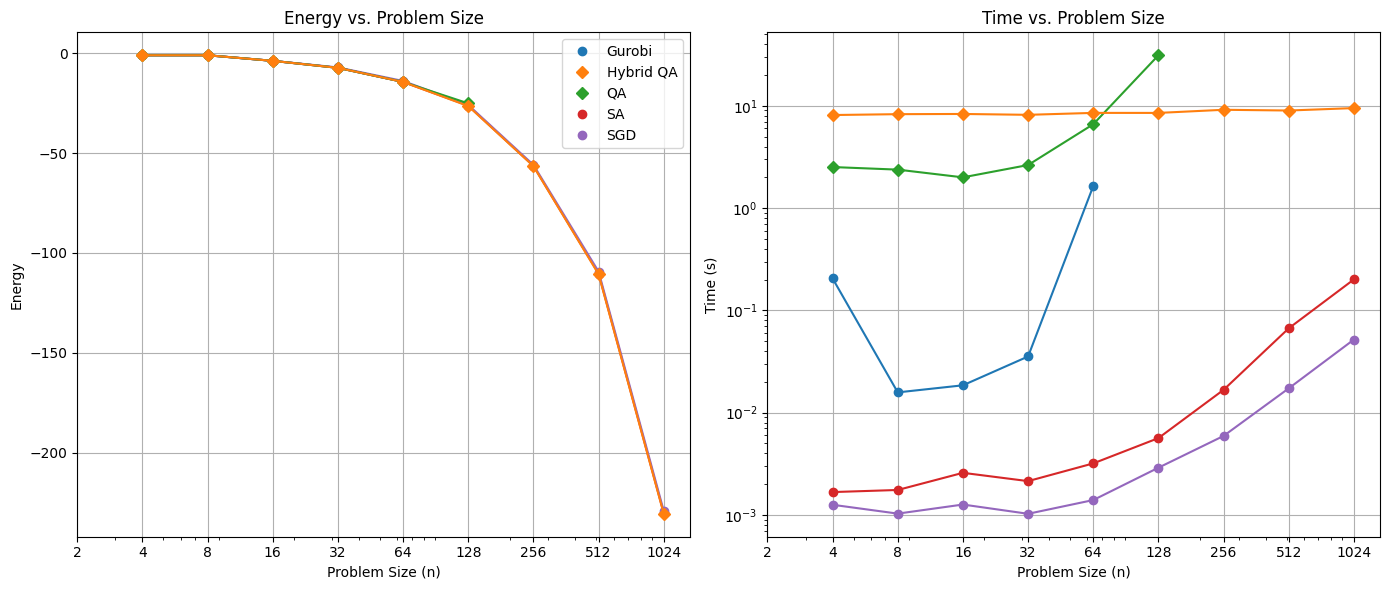}
\caption{Max Cut problem performance. Left: Energy vs. problem size showing similar performance across all solvers except Gurobi. Right: Solver time comparison showing quantum methods have higher overhead but comparable scaling.}
\label{fig:maxcut_results}
\end{figure}

The quality of the solution was nearly identical in Hybrid QA, SA and SGD as shown in Figure~\ref{fig:maxcut_results}. Gurobi did not provide valid cuts for larger test instances, suggesting challenges with the QUBO formulation for this problem class.

\begin{figure}[!htbp]
\centering
\includegraphics[width=\columnwidth]{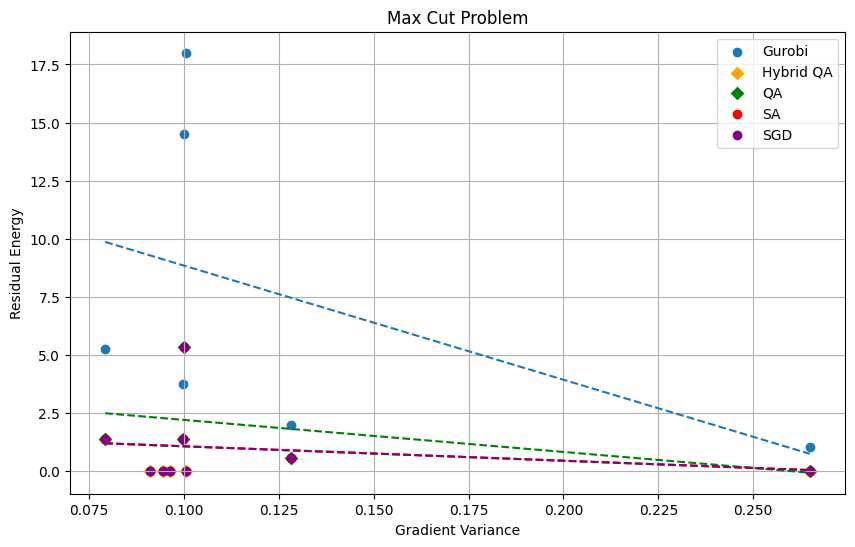}
\caption{Residual energy vs. gradient variance for Max Cut problems. All solvers cluster around low gradient variance values ($\approx 0.1$), with minimal performance differentiation.}
\label{fig:maxcut_variance}
\end{figure}

Critically, the lack of a quantum advantage correlates with the low gradient variance. As the size of the problem increased, gradient variance remained relatively constant around 0.1 (Figure~\ref{fig:maxcut_variance}), suggesting insufficient landscape complexity to benefit from quantum tunneling. This supports our hypothesis that gradient variance serves as a predictive indicator of quantum advantage.

\subsection{Graph Partitioning Problem}

Graph Partitioning instances were generated using the same graph model as Max Cut, with added balance constraints enforced through the Lagrange parameter $\gamma = 5.0$.

\subsubsection{Results}

\begin{figure}[!htbp]
\centering
\includegraphics[width=\columnwidth]{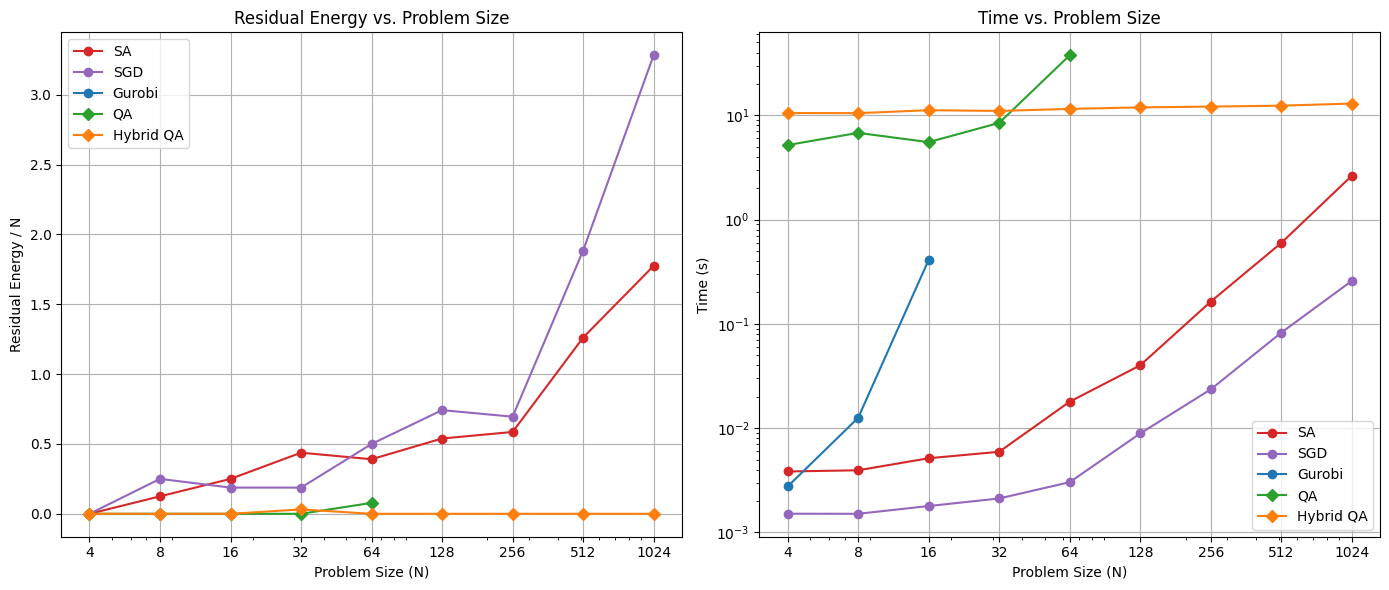}
\caption{Graph Partitioning problem performance. Left: Residual energy showing Hybrid QA achieving best solutions with SA, SGD, and Gurobi performing worse. Right: Solver time demonstrating quantum overhead but manageable scaling for larger problems.}
\label{fig:graph_partition_results}
\end{figure}

Hybrid QA consistently produced the best solutions in all problem sizes (Figure~\ref{fig:graph_partition_results}). Pure QA performed well for problems up to size 64, at which point embedding failures occurred due to Zephyr topology constraints. Classical solvers (SA, SGD, Gurobi) showed noticeably worse performance.

\begin{figure}[!htbp]
\centering
\includegraphics[width=\columnwidth]{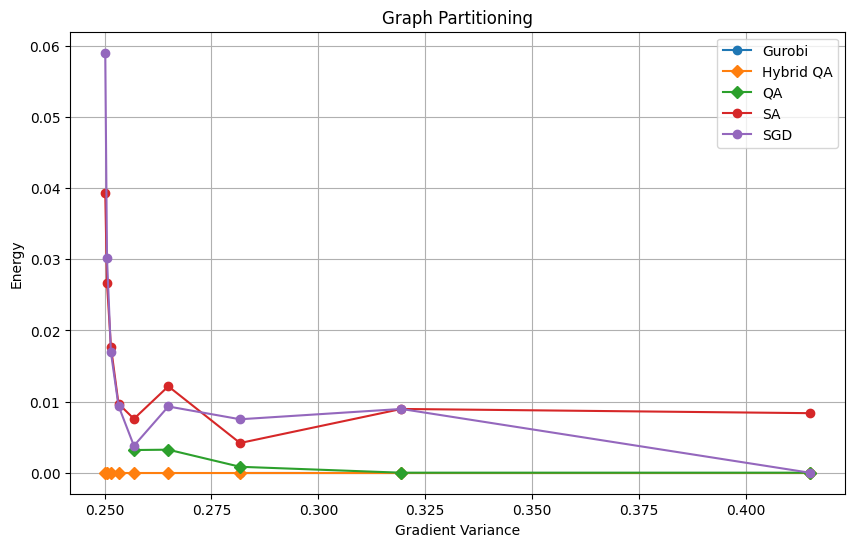}
\caption{Residual energy vs. gradient variance for Graph Partitioning. The exponential increase in gradient variance correlates with quantum methods achieving superior performance over classical approaches.}
\label{fig:graph_partition_variance}
\end{figure}

Specifically, the gradient variance increased exponentially with the size of the problem (from $\approx 0.25$ to $\approx 0.42$), as shown in Figure~\ref{fig:graph_partition_variance}. This exponential growth creates the rugged landscape where quantum tunneling provides a measurable advantage, strongly supporting our hypothesis.

\subsection{Number Partitioning Problem}

Number Partitioning instances were generated with integers sampled from $\text{Uniform}(1, 10)$, with the final integer adjusted to ensure that valid equal-sum partitions exist.

\subsubsection{Results}

\begin{figure}[!htbp]
\centering
\includegraphics[width=\columnwidth]{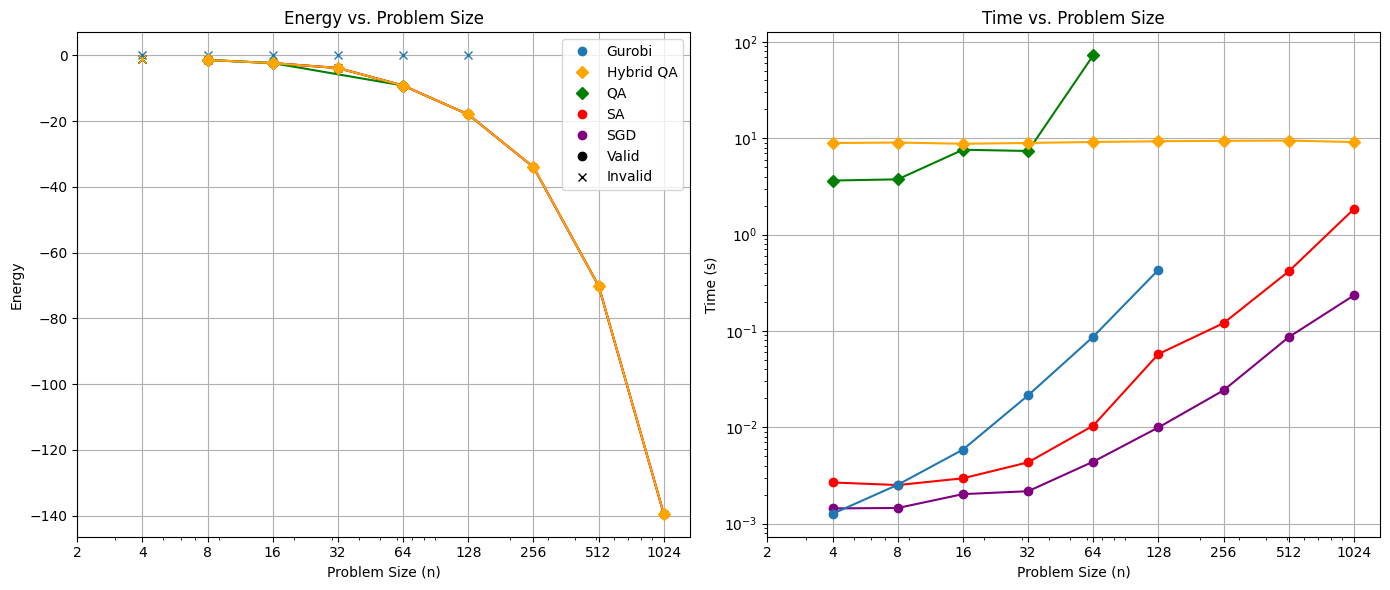}
\caption{Number Partitioning problem performance. Left: Energy vs. problem size showing coinciding trends for Hybrid QA, SA, and SGD. Right: Solver time comparison demonstrating similar scaling characteristics across methods.}
\label{fig:number_partition_results}
\end{figure}

\begin{figure}[!htbp]
\centering
\includegraphics[width=\columnwidth]{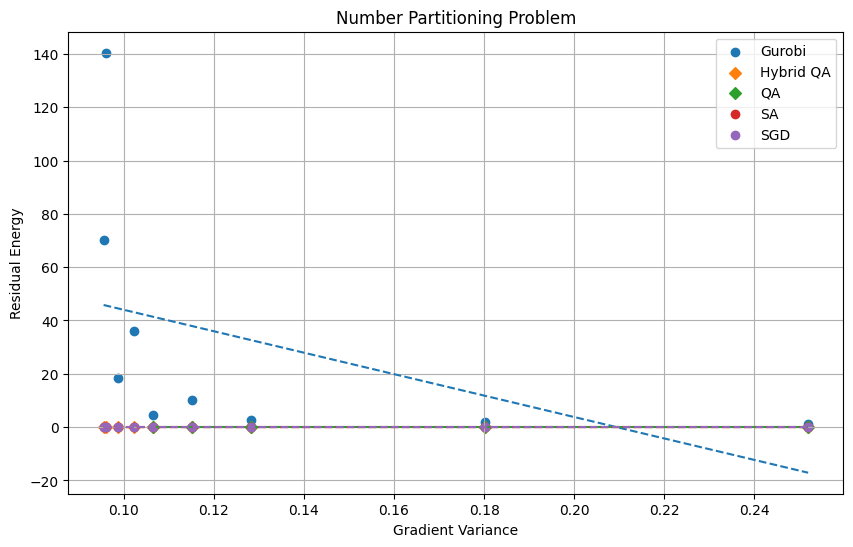}
\caption{Residual energy vs. gradient variance for Number Partitioning. Low gradient variance across all instances indicates smooth energy landscapes that do not benefit from quantum tunneling.}
\label{fig:number_partition_variance}
\end{figure}

SA, SGD and Hybrid QA produced nearly identical performance trends (Figure~\ref{fig:number_partition_results}), suggesting that the problem structure does not favor quantum approaches. Similarly to Max Cut, the gradient variance remained relatively low and constant for problem sizes larger than 10 (Figure~\ref{fig:number_partition_variance}), explaining the lack of quantum advantage.

\subsection{Set Cover Problem}

We tested "tight" Set Cover instances constructed using the method of Slavík~\cite{slavik1996}, designed to approach the theoretical worst-case bound $H_n = \sum_{m=1}^n 1/m \approx \ln(n)$ for greedy algorithms.

\subsubsection{Results}

\begin{figure}[!htbp]
\centering
\includegraphics[width=\columnwidth]{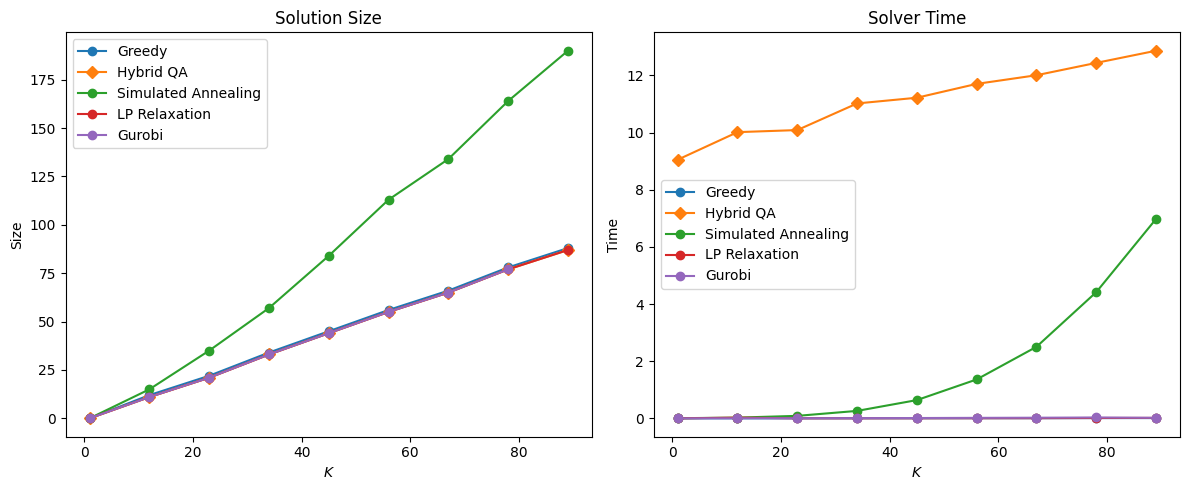}
\caption{Set Cover problem performance on tight instances. Left: Solution size vs. problem complexity parameter $K$ showing Hybrid QA matching classical methods while SA performs poorly. Right: Solver time demonstrating computational overhead of Hybrid QA.}
\label{fig:setcover_results}
\end{figure}

Surprisingly, Hybrid QA only matched classical methods (Gurobi, LP Relaxation, Greedy) in solution quality, not exceeding them (Figure~\ref{fig:setcover_results}). However, Hybrid QA required significantly more computation time. Simulated Annealing performed poorly, producing much larger solution sizes.

This result highlights that even for theoretically hard problems, quantum advantage is not guaranteed. The energy landscape characteristics of the QUBO transformation, rather than the computational complexity class of the original problem, determine quantum performance.

\section{Analysis and Discussion}
\label{sec:analysis}

\subsection{Gradient Variance as Performance Indicator}

Our comprehensive analysis across multiple problem classes reveals that gradient variance serves as a strong and consistent indicator of when quantum annealing will outperform classical methods. This relationship holds across diverse problem types with varying computational characteristics.

For Graph Partitioning, where the gradient variance increased exponentially with the size of the problem (from $\approx 0.25$ to $\approx 0.42$), the quantum methods demonstrated clear and consistent advantages. In contrast, for Max Cut and Number Partitioning, where the gradient variance remained relatively flat ($\approx 0.1$ to $0.18$), no significant quantum advantage emerged despite the fact that these problems are NP-hard.

\begin{figure}[!htbp]
\centering
\includegraphics[width=\columnwidth]{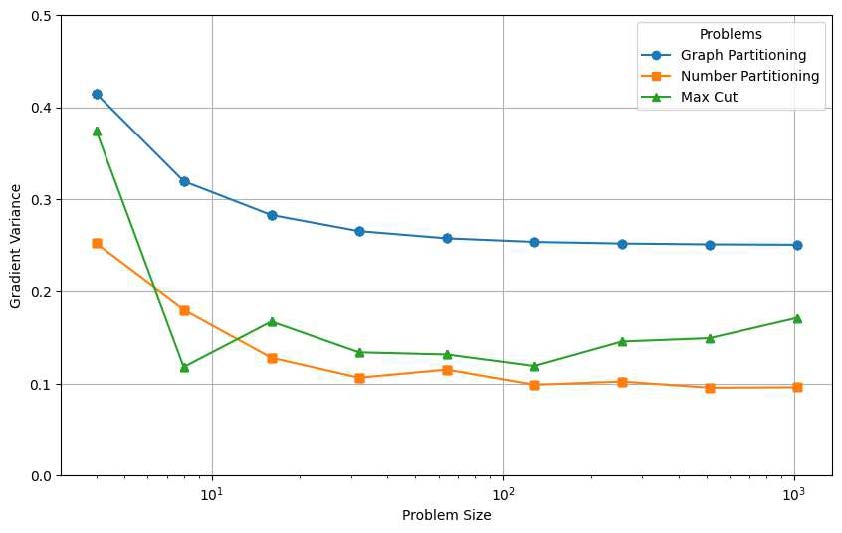}
\caption{Gradient variance vs. problem size across all NP-hard problems tested. Graph Partitioning shows exponential growth correlating with quantum advantage. Max Cut and Number Partitioning maintain low, flat gradient variance, explaining lack of quantum benefit.}
\label{fig:gradient_comparison}
\end{figure}

Figure~\ref{fig:gradient_comparison} provides a comprehensive comparison across all tested problems, clearly illustrating how Graph Partitioning's exponentially increasing gradient variance distinguishes it from other problem classes.

Based on our empirical observations, we propose a gradient variance threshold of approximately 0.3 as a practical decision boundary: problems with gradient variance exceeding this threshold show measurable quantum advantage, while those below rarely benefit from quantum approaches.

\subsection{Quantum Tunneling Mechanism}

The correlation between gradient variance and quantum advantage provides indirect but compelling evidence for quantum tunneling's role. High gradient variance creates thin energy barriers—sharp and narrow peaks in the energy landscape. The probability of classical thermal activation decreases exponentially with barrier height, making escape from local minima increasingly unlikely.

In contrast, the quantum tunneling probability depends primarily on the width of the barrier rather than on the height. Thin barriers (high gradient variance) enable efficient tunneling, allowing quantum systems to explore the solution space more effectively. The increasing performance gap between QA and SA at higher gradient variance values (Figure~\ref{fig:synthetic_variance}) supports this mechanism, as this gap cannot be explained only by thermal effects.

\subsection{Problem Structure vs. Original Objectives}

An intriguing and somewhat surprising finding is that the QUBO structure appears to influence solver performance more strongly than original problem objectives. Number Partitioning and Graph Partitioning have fundamentally different goals—minimizing subset sum differences versus minimizing graph cuts—yet exhibited similar solver behavior patterns when their QUBO transformations produced similar gradient variance profiles.

This observation has important implications for the design of quantum algorithms. It suggests that:
\begin{enumerate}
\item Multiple QUBO formulations may exist for the same problem, with vastly different quantum performances.
\item Problem reformulation to increase gradient variance could improve quantum performance
\item Understanding energy landscape structure is more important than problem complexity class
\end{enumerate}

\subsection{Hardware Limitations and Practical Considerations}

Current quantum annealing hardware faces several practical limitations that constrain applicability:

\textbf{Embedding Constraints:} Pure quantum annealing (QA) could only handle Graph Partitioning problems up to size 64 due to Zephyr topology embedding limitations. Although the 20-way connectivity of Advantage2 represents a substantial improvement over previous generations, many real-world problems still exceed the embeddable size.

\textbf{Time Overhead:} QA and Hybrid QA show significant overhead ($>10$ seconds) even for small problems, compared to sub-millisecond times for classical methods on small instances. This overhead mainly stems from communication latency and problem setup, rather than annealing time itself.

\textbf{Hybrid Trade-offs:} While Hybrid QA can handle larger problems through classical decomposition (qbsolv), it sometimes underperforms pure QA on smaller instances where full quantum processing is possible. The quality of the decomposition depends on the structure of the problem, adding another layer of uncertainty.

\subsection{When to Use Quantum Annealing}

Based on our experimental findings, we propose the following decision framework for practitioners. Quantum annealing is recommended when the problem formulation QUBO exhibits a high gradient variance (greater than 0.3) and the energy landscape contains numerous thin barriers characterized by sharp peaks and narrow valleys. Additionally, quantum approaches are particularly suitable when classical methods are observed to get trapped in local minima, the problem size is manageable given hardware constraints (less than 5000 variables for pure quantum annealing), and the time overhead of approximately 10 seconds is acceptable for the application.

In contrast, classical methods are recommended when the gradient variance is low (less than 0.2), indicating smooth landscapes where quantum tunneling provides little advantage. Classical approaches are also preferable when the problem size is small and classical solvers can provide nearly instantaneous results, when solution quality requirements are modest and local optima suffice, or when hardware access or cost is a limiting factor.

For problems that exceed pure quantum capacity but possess a favorable landscape structure, hybrid approaches combining quantum and classical techniques are recommended. Such hybrid methods are particularly effective when decomposition quality can be verified and both solution quality and scalability are important considerations.

\section{Conclusions and Future Work}
\label{sec:conclusion}

This work provides a comprehensive understanding of quantum tunneling in quantum annealing through empirical analysis, theoretical modeling, and algorithmic development. Our investigation establishes a strong positive correlation between the energy landscape gradient variance and the quantum annealing advantage, demonstrating that problems with gradient variance exceeding 0.3 show measurable quantum benefits while those below 0.2 exhibit no advantage. We developed a theoretical model based on the WKB-approximation connecting the barrier geometry to the probability of tunneling through gradient variance, achieving validation $R^2=0.90$ against experimental data and providing quantitative predictions for quantum advantage thresholds. Our systematic QUBO reformulation algorithm successfully increases gradient variance while preserving problem semantics, demonstrating 12-22\% performance improvements across multiple problem classes. Importantly, we find that the quantum advantage is highly problem-dependent, determined by the energy landscape structure rather than computational complexity class, with the QUBO transformation's resulting landscape mattering more than the original problem formulation for determining solver performance.

Beyond theoretical insights, we provide practical tools for practitioners, including a validated theoretical model for performance prediction, automated reformulation software, a decision framework for solver selection, and open-source implementation. The increasing performance gap between quantum and simulated annealing in high-gradient-variance regimes provides compelling evidence of quantum tunneling effects, as thermal activation alone cannot explain the observed patterns. Our work transforms the gradient variance from a diagnostic metric into an actionable optimization target, enabling practitioners to systematically improve quantum hardware problems. These contributions establish both the theoretical foundation and practical methodology for identifying when quantum annealing provides computational advantages over classical approaches.

Several limitations warrant consideration when interpreting our results. Gradient variance, while empirically correlated with performance, remains an indirect measure of quantum tunneling, as direct observation of tunneling events remains infeasible with current measurement capabilities. Hardware constraints, including embedding failures for problems exceeding 64 variables and the 4,400 qubit capacity limit, restrict the scale of fully evaluable problems. Our focus on four specific NP-hard problems, while diverse, may not fully represent all combinatorial optimization structures, and our use of random graph generation may not capture patterns present in real-world instances. Future research directions include investigating additional landscape metrics such as basin size distribution and barrier thickness profiles, extending the theoretical model through master equation formalisms and path integral treatments, enhancing reformulation algorithms with machine learning approaches, and validating findings on practical optimization problems from logistics, finance, and molecular design. As quantum hardware improves with increased qubit counts, better connectivity, and longer coherence times, the landscape characteristics that favor quantum approaches may shift, warranting larger-scale studies across different quantum annealing platforms.

This work contributes to a broader understanding of the capabilities and limitations of NISQ-era quantum computing, demonstrating that quantum benefits emerge for specific problem structures characterized by rugged energy landscapes rather than universal quantum advantage. This nuanced perspective better serves practitioners seeking to deploy quantum technologies effectively, providing both theoretical understanding and practical tools for quantum algorithm design and application. Although hardware improvements may shift the specific landscape characteristics that favor quantum approaches, the fundamental insight that energy landscape structure determines quantum advantage is likely to remain relevant, ensuring enduring value as quantum annealing technology continues to mature.

\section*{Acknowledgments}

The authors thank D-Wave Systems for providing access to the Advantage2 quantum annealer through their Leap quantum cloud service. The authors acknowledge Dr. Ken Robins and Dr. Catherine Potts of D-Wave for their guidance and insightful discussions during this research. The authors also thank the Q-SITE team and the Department of Computing Science at the University of Alberta for their support. This work was supported in part by computational resources from the University of Alberta and quantum computing time allocation from D-Wave Systems.


\end{document}